# Probing the pH Microenvironment of Mesenchymal Stromal Cell Cultures on Additive-Manufactured Scaffolds

*Ivan Lorenzo Moldero, Anil Chandra, Marta Cavo, Carlos Mota, Dimitrios Kapsokalyvas, Giuseppe Gigli, Lorenzo Moroni,\* and Loretta L. del Mercato\**

**Despite numerous advances in the field of tissue engineering and regenerative medicine, monitoring the formation of tissue regeneration and its metabolic variations during culture is still a challenge and mostly limited to bulk volumetric assays. Here, a simple method of adding capsules-based optical sensors in cell-seeded 3D scaffolds is presented and the potential of these sensors to monitor the pH changes in space and time during cell growth is demonstrated. It is shown that the pH decreased over time in the 3D scaffolds, with a more prominent decrease at the edges of the scaffolds. Moreover, the pH change is higher in 3D scaffolds compared to monolayered 2D cell cultures. The results suggest that this system, composed by capsules-based optical sensors and 3D scaffolds with predefined geometry and pore architecture network, can be a suitable platform for monitoring pH variations during 3D cell growth and tissue formation. This is particularly relevant for the investigation of 3D cellular microenvironment alterations occurring both during physiological processes, such as tissue regeneration, and pathological processes, such as cancer evolution.**

## 1. Introduction

Several strategies of increasing complexity have emerged to provide more and more biomimetic and efficient 3D scaffolds to be used as substrates for cell culture and tissue regeneration: far from being passive components, nowadays scaffolds play a significant role in tissue regeneration, balancing mechanical function, and mass transfer properties.[1–3] Methods to develop scaffolds range from manual techniques, such as particulate leaching[4,5] and freeze-drying,[6] that allow controlling architectural features[7,8] but are highly susceptible to external conditions, to more advanced processes in which scaffolds can be built layer by layer in an additive manner directly from computer data such as computer-aided design (CAD).[9,10] This last group of techniques is commonly known as additive manufacturing (AM) and includes stereolithography, fused deposition modeling (FDM), 3D inkjet printing, and selective laser sintering; the precise control over scaffold porosity, pore size, and interconnectivity obtained with AM techniques has shown a great influence on cellular behavior, leading to promising results in terms of tissue regeneration.[11–14]

Moving from 2D to 3D cultures has required a considerable effort in adapting the protocols used in 2D in order to understand more about how the 3D cell culture model affects cellular behavior and to allow direct comparison of the results.[15–17] In many cases, real-time monitoring methods have been developed by integrating biosensors into bioreactor systems. Nowadays this allows detecting during culture bulk volumetric variations in pH or oxygen consumption, and other metabolites production.[16,18–20] However, the majority of the assays provide bulk measurements that provide an overview picture of cell conditions within the 3D constructs without discriminating zonal variations and without detecting the gradients of gases, growth factors, and metabolites that are naturally present between the center and the periphery of an engineered tissue construct. In particular, monitoring pH during cell culture is essential for the maintenance of cellular viability and for improving tissue functions.[21] In tissue regeneration processes, the acid–base balance is fundamental for the remodeling process: in the specific case of bone, for example, metabolic acidosis causes calcium efflux,[22,23] while alkalosis decreases bone calcium efflux.[24,25]

Dr. I. L. Moldero, Dr. C. Mota, Prof. L. Moroni
Department of Complex Tissue Regeneration
MERLN Institute for Technology-Inspired Regenerative Medicine
Maastricht University
Maastricht 6229ER, The Netherlands
E-mail: l.moroni@maastrichtuniversity.nl
Dr. A. Chandra, Dr. M. Cavo, Prof. G. Gigli, Prof. L. Moroni,
Dr. L. L. del Mercato
Institute of Nanotechnology
National Research Council (CNR-NANOTEC)
Campus Ecotekne, via Monteroni, Lecce 73100, Italy
E-mail: loretta.delmercato@nanotec.cnr.it
Dr. D. Kapsokalyvas
Department of Molecular Cell Biology
Maastricht University Medical Center
UNS 50, Maastricht 6229ER, The Netherlands
Prof. G. Gigli
Department of Mathematics and Physics "Ennio De Giorgi"
University of Salento
via Arnesano, Lecce 73100, Italy

Consequently, extracellular acidosis promotes resorption, whereas alkalosis promotes formation/mineralization.[26–28] In cartilage regeneration it was found that at pH 7.2 bovine chondrocytes produce significantly more glycosaminoglycans (GAGs) than at pH 7.4, while at pH 6.4 the production of GAGs is only 10% of that produced at pH 7.2.[29,30] Therefore, tools to sense pH are necessary in order to accurately monitor pH levels and possibly find ways to regulate it.

Among the different methods available to monitor pH values within cell cultures, traditional techniques (such as pH-meter probes or microelectrodes) mainly provide an average value that does not take into consideration the heterogeneity of a 3D cell culture system. Newer advanced techniques consist of microneedle sensors that can penetrate the sample to be analyzed without causing significant damage to the sample. However, their application is generally limited to analysis in bulk structures such as brain fluid, intestine, bladder, and skin.[31–34] These systems require big volumes of culture medium and often compromise the sterility of the experiment.[35]

To circumvent these problems, fluorescent sensors have been developed to study the metabolic state at a cell size scale and to provide a real-time detection of acidification basing on the changes of fluorescence intensity induced by pH variations.[36,37] These systems represent a minimally invasive sensing technology, since they can be used without removing or damaging cells during culture.[38] Among carbon-based fluorescent nanomaterials, pH sensors derived from graphene quantum dots[39–42] and carbon dots are common.[43–46] Their exquisite fluorescence properties make them very advantageous in many aspects, however they have a tendency to interact with cellular mechanisms and thus can interfere with the studies in question.

To precisely sense the biological signals in a cellular microenvironment, a probe with microdimensions that can individually register the pH is desirable: for this purpose, sensors with microscale size have been developed for effective biosensing,[47] The choice of fluorescent microparticles-based sensors has an advantage over fluorescent nanoparticles-based sensors during fluorescence microscopy analyses. Namely, nanoparticles due to their small size are difficult to be resolved by fluorescence microscopy as individual particles.[48] Therefore, it is practically impossible to monitor individual nanoparticles-based sensors in space and time. Microparticles-based sensors on the other hand can be visualized clearly by fluorescence microscopy and their intensity variation can be continuously tracked in a spaciotemporal manner. Further, it is also important to consider the position of the sensors, as sensing of local pH in close vicinity of cells by multiple small sensors is more informative compared to an averaged pH readout by a distant bulk pH sensor that cannot register smaller pH gradients that exist spatially around the cells.[49] Undeniably, the combination of several technologies, such as AM for the control of scaffolds' porosity and fluorescent sensors for the monitoring of local pH changes, offers great promise for improving tissue regeneration efficiency and accelerating the consequent translation from bench to bedside in the near future.

The aim of this work was to present a combined system that lays the basis for a new generation of smart sensing scaffolds from their advanced fabrication strategies to their performance in vitro. We present all the intermediate steps necessary to reach this goal, namely (i) the development and characterization of fluorescent pH-sensing capsules based on seminaphtharhodafluor (SNARF-1), a ratiometric pH indicator probe with pKa of ≈7.5 and appreciable variation in fluorescence emission ratio against respective pH values,[50,51] (ii) the development of 3D scaffolds with defined geometry obtained by FDM, and finally (iii) the validation of the integrated pH-sensing capsules/FDM scaffolds system by calibrating, monitoring, and mapping pH variations of seeded human mesenchymal stromal cells (hMSCs) in different areas of the constructs compared to 2D conditions.

## 2. Results and Discussion

### 2.1. Properties of Capsules-Based pH-Sensors

The pH response of capsules-based pH sensors was measured by incubating the capsules in buffer solutions at known pH values and by recording the fluorescence emission ratio ($I_{595\,nm}/I_{640\,nm}$) under 543 nm excitation wavelength. **Figure 1**a shows three representative images of capsules under pH 5.64, 6.58, and 8.28, respectively. In Figure 1b, the intensity ratio ($I_{595}/I_{640}$) recorded by the CLSM analysis of capsules incubated in media at different pHs from 5.64 to 8.28 is reported.

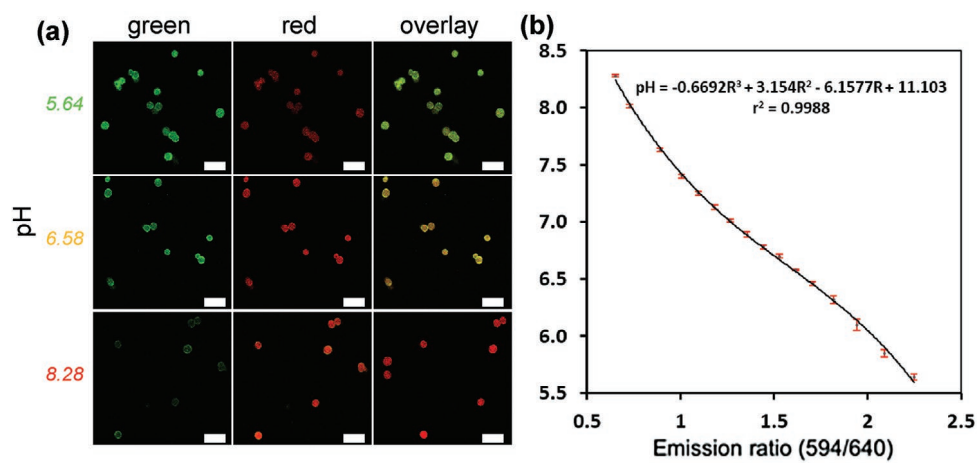

**Figure 1.** Calibration of capsules-based pH sensors. a) Representative CLSM micrographs showing the pH dependence of capsule's fluorescence in pH-adjusted cell media. Green channel (false color, 580–610 nm), red channel (625–655 nm), and overlay of the fluorescence channels are reported ($\lambda_{exc}$ = 543 nm). Scale bars: 10 μm. b) Ratiometric calibration curve of capsules-based pH sensors on fluorescence intensity ratio of green (false color) and red channels derived from CSLM micrographs. Tested pH values from 5.64 to 8.28.

In accordance with the photophysical properties of the pH indicator dye SNARF,[51] the emission of the capsules strongly depends on the local pH, that is, at acidic pH the capsules predominantly emit in the green (false color), whereas at basic pH they emit in the red region of the visible spectrum. Figure 1b shows a plot of the fluorescence intensity ratio ($I_{595}/I_{640}$) as a function of pH obtained by quantifying the fluorescence intensity of the capsules at different pH values. The performed analysis indicated a direct relationship between SNARF fluorescence and proton concentration (fit parameter values are shown in the corresponding graph). The equation of the fit calibration curve was used to estimate an unknown pH experienced by the capsules in a region of interest. These data are in agreement with our previous results, in which we studied the fluorescence response of SNARF–dextran conjugates loaded in the cavities of polyelectrolyte capsules inside living cells.[52] The pKa calculated for SNARF-1-dextran conjugate was 7.09 ± 0.04, which is lower compared to the pKa of SNARF-1 (pKa ≈7.5). The possible reason for lowering of pKa could be the result of interaction between SNARF-1 and surrounding microparticle matrix.[55] Here, it is to be noted that the possibility of change in emission ratio for a specific pH due to photobleaching is very unlikely as both the emissions have similar quenching behavior and thus the ratio is always conserved. Additionally, lower exposure time during microscopy rather than time-lapse imaging would further reduce the extent of photobleaching that is generally observed at alkaline pH values.[55,56]

After successfully analyzing the pH sensitivity of the capsules, it was important to study their effectiveness of pH sensing and localization in a cellular environment. As the capsules are of a micrometer dimension they can be uptaken from the cells by endocytosis.[57,58] Therefore, we studied the extent of uptake of the pH-sensor capsules by incubating the sensors with hMSCs for 7 d. It was observed that the cells can uptake up to 25 ± 4% of the capsules added to the cell culture media, while leaving rest of the capsules outside (**Figure 2**a). The intracellular and extracellular pH-sensor capsules can be easily differentiated by observing their ratiometric fluorescence emission using fluorescence microscopy. Internalized SNARF-based pH-sensor capsules showed higher green fluorescence emission intensity due to acidic intracellular pH, whereas extracellular capsules exhibited main orange-red emission due to comparatively higher pH of the surrounding medium. Therefore, by merging the fluorescence microscopy images of the capsules acquired using green and red emission filter channels, the position of the capsules with respect to the cell can be identified very easily (Figure 2a).

Additional information about the intracellular location of the capsules was obtained by incubating ad hoc made FITC-based capsules with hMSCs treated with lysotracker, which stains lysosomes (Figure 2b). As expected, FITC-based capsules located in the extracellular medium exhibited prominent green emission due to higher extracellular pH of the cell culture medium (solid arrows in Figure 2b), whereas capsules internalized by cells showed strong colocalization with red-stained acidic lysosomes (dashed arrows in Figure 2b) as result of the well-known quenching of FITC emission at low pH. Furthermore, the internalized capsules displayed the typical deformation (dashed box in Figure 2b), which is known to be due to the mechanical pressure within the intracellular vesicles that leads to squeezing of the capsules.[59] Namely, this deformation does not impair the sensing properties of the capsules that retain both the pH-sensitive and the reference fluorophores within their cavities, but additionally confirm their intracellular localization.[52] Figure S4, Supporting Information shows the normal lysosome quantity and distribution in hMSCs growing in 3D AM scaffolds, where cells were separately stained with calcein and lysotracker (control). The internalization of capsules at day 7 was further assessed via TEM analysis (Figure 2c,d). TEM images indicate a homogeneous cytoplasm, showing vesicular structures, most likely mitochondria, lysosomes, and endosomes. Several capsules can be seen inside the cells, some exhibiting partial deformation as a result of the internalization process (solid arrows in Figure 2c,d).

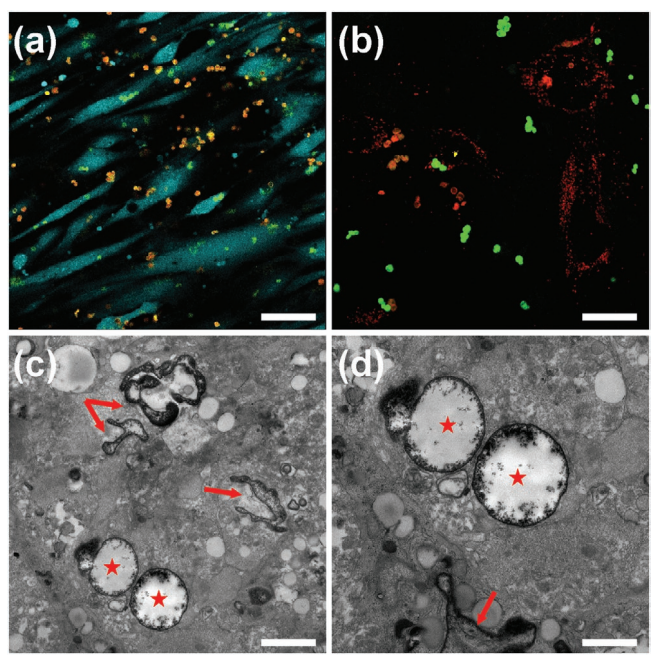

**Figure 2.** Internalization of capsules by hMSCs cultured in additive manufacturing scaffolds. a) Representative CLSM image of hMSCs incubated with pH-sensing ratiometric SNARF-based capsules, with orange-red capsules indicating extracellular location and green capsules indicating intracellular location. The cell cytoplasm is stained using calcein (pseudo-cultured in cyan). b) Representative CLSM image of hMSCs incubated with pH-sensing FITC-based capsules, with green capsules indicating extracellular location and red capsules indicating colocalization with lysosomes stained with lysotracker (red emission signal). Dotted arrows and squares indicate examples of intracellular located capsules, straight lines indicate examples of extracellular located capsules. Only overlay images of merged channels are shown for the sake of clarity. Scale bars: 40 µm. c,d) Representative transmission electron microscopy (TEM) images showing the uptake of the capsules by hMSCs. Stars, examples of light-dense capsules, while arrows indicate deformed capsules. Scale bars: 2 µm. CLSM and TEM images were taken after 7 d of cell–capsules interaction.

### 2.2. Scaffold Toxicity

After thorough investigation of the capsules for their pH-sensing capabilities our next goal was to study the suitability of the additive manufactured scaffolds for culturing hMSCs and mapping pH changes within it. The growth of cells in the scaffolds typically starts with cell seeding, followed by adhesion and proliferation. As the cells divided over time and started to occupy the empty pores in the scaffolds, they disposed themselves in a spiral-like shape, which is typical of hMSCs growth in these scaffolds as previously observed.[60,61] At instances, where the pore is in the process of filling, it will be referred to as a filling pore. A filled pore on the other hand will denote a pore completely filled with cells. The fibrous boundary of the additive manufactured scaffolds is denoted by fiber. **Figure 3**a indicates the different regions that are typically formed during growth of cells on the scaffold. The assessment of cell viability on the scaffold was done by using a live-dead assay, where calcein-AM and ethidium homodimer-1 were used to stain live and dead cells, respectively.

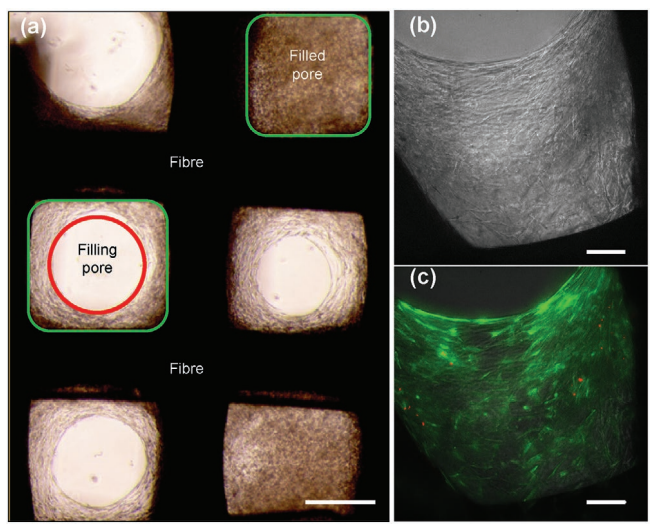

**Figure 3.** Cell growth in the additive manufactured porous scaffolds. a) Brightfield microscopy image showing hMSCs growing on the additive manufactured scaffold. Filling pore in the image represents volume yet to be occupied by the proliferating cells. A completely filled pore is denoted as "filled pore." Pore size of scaffolds is approximately 450 × 450 µm in the x-y plane, and 170 µm in the z plane. Scale bar: 250 µm b,c) Live-dead staining assay of the hMSCs proliferating in the scaffold pores (day 7) to detect viable (green) and dead (red) cells, respectively. (b) Bright-field, (c) overlay of green (calcein) and red (EthD-1) channels. Scale bars: 100 µm.

As can be observed in Figure 3c, live-dead assay showed almost all cells in the pore's scaffold stained green, indicating high viability of hMSCs. Cells appeared with a spindle-like shape embedded in their own extracellular matrix (ECM) bridging the pore space between two fibers. To further investigate the suitability of the 3D scaffolds in supporting the hMSCs growth, imaging of collagen produced by hMSCs was performed. Collagen formation was observed through fluorescent labeling with CNA-35 (Figure S5, Supporting Information). Collagen was well developed and had the typical fibrillar form as observed with second harmonic generation (SHG) microscopy (Figure S6, Supporting Information). SHG can be observed only in fibrillar collagen, usually Type I or III.[62] As can be observed in Figures S5 and S6, Supporting Information, collagen fibers can be visualized, which emphasizes their normal growth and viability.

## 2.3. Visualization of pH Microenvironment

After pH calibration of the capsules, they were used to image the local pH microenvironment of hMSCs growing on the 3D scaffolds. In **Figure 4** and Figure S7, Supporting Information images of cells growing on the scaffolds, in either static or dynamic conditions, with pH-sensing capsules are presented. In Figure 4a,b the intensity-based images are introduced. Cytoplasm staining with calcein is coded with magenta and capsules staining is coded in the green ($I_{595}$) and red ($I_{640}$) channels. In Figure 4c,d coding of cytoplasm is converted to a gray scale, and the red, green, and blue channels code for different ranges of pH. Based on the ratio of $I_{595}$ and $I_{640\ nm}$ and according to pH calibration formula, pH values from 5.0 to 6.4 are coded in red, from 6.4 to 6.8 are coded in green, and from 6.8 to 9.0 are coded in blue. In these images, clear visualization of capsules with different pH values is possible. Red capsules are mainly inside acidic compartments of the cell such as lysosomes, while blue capsules are in the extracellular space. When comparing dynamic with static cultures, capsules were more homogeneously distributed and more abundantly disperse in the engineered tissue (Figure S7, Supporting Information).

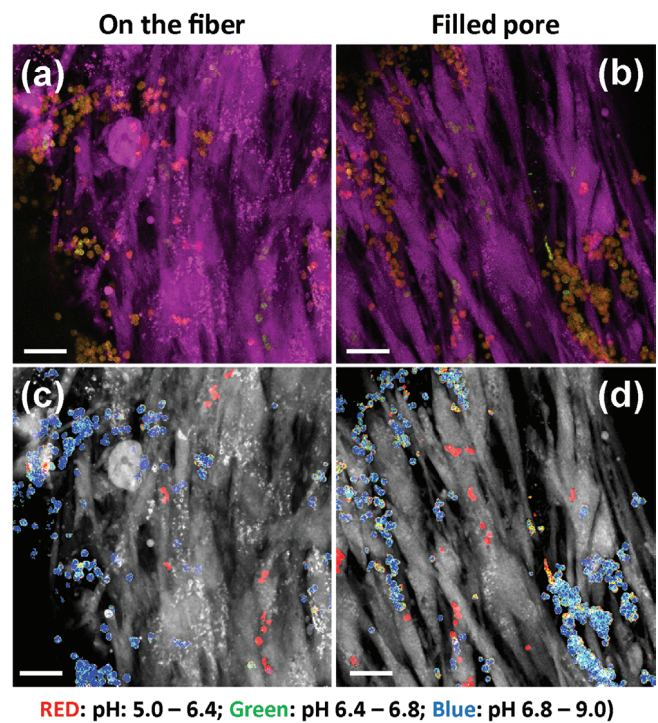

**Figure 4.** Maximum intensity z-projection images representing hMSCs and pH-sensing capsules. a,c) Capsules sensing pH around cells growing on the fiber and b,d) capsules sensing pH around cells in the filled pore of the 3D scaffold (day 7). (a, b) Intensity-based images, Magenta: Cytoplasm (calcein), Green: $I_{595}$ and Red: $I_{640}$. Green and red correspond to the signal of the pH-sensitive capsules. (c,d) pH images. In these images the cytoplasm coding is converted to gray for better clarity and the red, green, and blue coding indicates different pH ranges according to the ratio values of the $I_{595}$ and $I_{640}$ and the conversion to pH based on the calibration formula. pH coding, Red: 5.0–6.4, Green: 6.4–6.8, and Blue: 6.8–9.0. Scale bars: 20 µm. Field of view in (a) 148 × 148 µm² max projection of 25 sections (z step 2 µm); in (b) 147 × 147 µm² max projection of 11 sections (z step 2 µm).

## 2.4. Quantification of pH Microenvironment Changes in Engineered Tissue

The immediate pH in local microenvironment of a growing cell can have a huge impact on its growth and metabolism. As hMSCs are known to change their properties significantly due to change in pH,[63] using the pH-sensing capsules we therefore studied the pH at different spatial locations in 3D and considering the additive manufactured scaffolds as a geometrically defined 3D cell culture system where pH mapping could be more easily addressed. A representation of pore position in the scaffold is described in Figure S7, Supporting Information. As the cells can either grow on the fiber of the scaffold or in the pore region, we estimated pH in different regions using the pH-sensing capsules. This is particularly relevant, since it allows generating a spatial mapping of pH within a 3D cell culture system, which has been unreported in the literature to the best of our knowledge. Indeed, mapping of pH in 3D conditions is still in its embryonic phase, and the only work proposing it does not present a mixed population of cells and sensors, but pH-sensing substrates are put in contact with 2D culture platforms,[49] thus reducing the study to a 2D mapping.

The fluorescence microscopy results indicate that pH at all the locations was changing with time, where in general the pH reached more acidic values inside the scaffolds compared to 2D culture (**Figure 5**a,b). The pH decrease was smaller on the scaffold fibers in comparison to the scaffold pores. Limited growth of the cells on the fiber due to lack of neighboring cells could be a possible reason for less acidic pH at the fiber, since the total amount of acidic metabolites released depends on the cell proliferative state. Pores in general exhibited a relatively more acidic microenvironment, but we also wanted to explore the effect of location of the pore in the scaffolds. Dimension and complexity of the scaffolds can have a deep effect on diffusion of media, metabolites, and movement of cells across them. In this study, it was observed that pores on the corner of the scaffold were most acidic with average estimated pH of 6.99 ± 0.07 after 7 d followed by pH 7.08 ± 0.14 and pH 7.15 ± 0.15, respectively, for the pores on the edge and inside the scaffold. More acidic pH at the corner compared to edge and central regions of the scaffold seems a counterintuitive result, but could be explained by the fact that diffusion of gases occurs more easily close to the boundary and can cause more acidity due to $CO_2$-mediated pH control.

One more reason for interpore pH variation could be availability of more media at the edge causing enhanced growth of cells which eventually caused further decrease in pH at those locations after 7 d (Figure 5c,d). The variation in pH at the corner location due to maximum exposed surface area for diffusion compared to a lower diffusion at the edge and minimum diffusion in inner locations of the scaffold started becoming evident just in 2 d after seeding of cells in the scaffold (Figure S8, Supporting Information). This early generation of pH gradient as observed by the pH microsensors emphasizes the importance of the geometry and architecture of the scaffold in providing a defined microenvironment that has the potential to drive cell fate in terms of growth, production of desired metabolites, or in controlling the process of differentiation.

Notably, alkaline pH values of the αMEM media on the scaffold without the cells were observed starting with day 1 till day 7. This is most likely due to Le Chatelier's principle, for which the pH of culture medium is known to increase when not mediated by the presence of $CO_2$, which is the case during 2-photon microscopy analyses. In presence of cells the alkaline species may have been neutralized by the growing cells, but without the cells the media remained alkaline. This control nonetheless demonstrates that the pH of this sample remained stable and the great pH changes observed in the scaffolds and the 2D cultures were caused by the cells induced acidity. Our standard differentiation protocols for the generation of tissue-like constructs begin after 7 d hMSCs culture was grown in additive manufactured scaffolds with proliferation media to allow for sufficient cell proliferation and a basal ECM formation. For this reason and the complexity of the assay, we have focused on this first 7-d period. Indeed, in future studies it will be important to follow up the measurement of pH microenvironment of hMSCs cultures on additive manufactured scaffolds for 3–4 weeks in differentiation medium, which are typical time points to evaluate cell differentiation and tissue formation. In the context of skeletal tissue formation, it would be interesting for example to study the impact of pH on local mineralization.

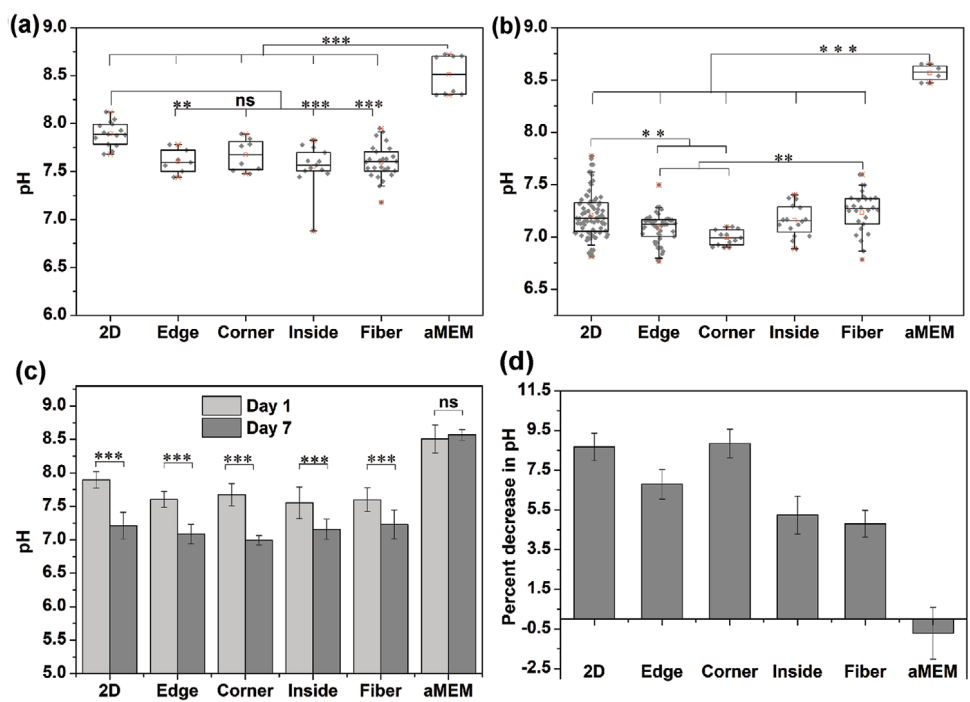

**Figure 5.** Extracellular pH estimation using pH-sensing capsules within different regions of the scaffold populated with hMSCs a) day 1 pH distribution and b) pH distribution at day 7. c) Plot showing side-by-side comparison of pH between day 1 and day 7. d) Percent reduction in extracellular pH at the end of 7 d compared to pH at day 1. Statistical information (*$p$-value < 0.05, **$p$-value < 0.01, ***$p$-value < 0.0001, ns: not significant).

## 3. Conclusions

In summary, we describe the fabrication and evaluation of a hybrid system composed of capsules-based optical sensors and 3D scaffold with predefined geometry to monitor local pH fluctuations during cell culture of hMSCs. By means of confocal fluorescence microscopy we were able to generate spatially resolved maps of pH across the 3D cultures of hMSCs-containing scaffolds. This system also allowed us to map the temporal evolution of pH across a 7-d period.

Altogether, the present model system demonstrates how 3D porous scaffolds can be investigated under confocal microscopy in real time while simultaneously detecting the pH variations occurring in the local microenvironment. Designing 3D-printed biomaterials that can provide more suitable conditions for the cells being cultured remains a challenge. We will continue to use this system to spatially map pH gradients in our 3D cultures and to determine how changes in culture conditions, such as cell type or co-culture of different cells, affect those gradients. The optimization of scaffolds design parameters based on this kind of microenvironment analysis is more intuitive and relevant. In applications such as tissue engineering and artificial organ development, one of the major problems is too steep physical and biochemical gradients. This simple method of assessing pH in the cellular microenvironment is thus one step forward in more advanced and smarter bio-scaffolds.

## 4. Experimental Section

*Chemicals*: Poly(sodium 4-styrenesulfonate) (PSS, ≈70.000 molecular weight (MW)), poly(allylamine hydrochloride) (PAH, ≈56.000 MW), calcium chloride dehydrate ($CaCl_2$, 147.01 MW), sodium carbonate ($Na_2CO_3$, 105.99 MW), ethylenediaminetetraacetic acid disodium salt dehydrate (EDTA), polylactic acid (85 000–160 000 MW), glutaraldehyde solution grade I (50% in $H_2O$), sodium cacodylate trihydrate, dichloromethane, and acetone were purchased from Sigma-Aldrich. Aminodextran (500 000 MW) and 5-(and-6-)-carboxy-seminaphthorhodafluor-1 acetoxymethyl ester acetate (SNARF-1, 567.5508 MW) were obtained from Invitrogen, fluorescein 5(6)-isothiocyanate (FITC, 389.38 MW) from Sigma. All chemicals were used as received. Ultrapure water with a resistance greater than 18.2 MΩ cm was used for all experiments.

*Synthesis of Capsules-Based pH Sensors*: The pH indicator dye SNARF-1 was conjugated to the nonfluorescent aminodextran and subsequently loaded inside porous calcium carbonate ($CaCO_3$) microparticles obtained via coprecipitation of equal volumes of $Na_2CO_3$ (0.33 M) and $CaCl_2$ (0.33 M) solutions.[21,52] As the next step, SNARF-1-dextran filled $CaCO_3$ microparticles (size distribution around 3.5–5 µm) were coated by multiple layer-by-layer assembly of the oppositely charged PSS (2 mg mL$^{-1}$, 0.5 M NaCl, pH = 6.5) and PAH (2 mg mL$^{-1}$, 0.5 M NaCl, pH = 6.5) polyelectrolytes. This procedure was repeated until six layers were assembled around the spherical microparticles, thus providing a multilayer shell (PSS/PAH)$_3$. In the last step, the sacrificial $CaCO_3$ templates were removed by complexation with EDTA buffer (0.2 M, pH 7). Finally, the multilayer polyelectrolyte capsules, carrying the SNARF-1-dextran conjugate in the cavities, were stored as suspension in 2 mL of Milli-Q water at 4 °C. After template removal the number of capsules per volume was determined by direct counting in a defined volume with a hemocytometer chamber under an inverted optical microscope. From one synthesis we obtained $9.47 \times 10^8$ capsules per milliliter.

For colocalization analyses, a batch of capsules loaded only with FITC-dextran was prepared by using the same procedure described here for synthesizing SNARF-based capsules but replacing SNARF-1-dextran conjugate by FITC-dextran conjugate.[21,53]

*Fabrication and Characterization of 3D Scaffolds*: Poly(ethylene oxide terephthalate) and poly(butylene terephthalate) random block copolymer (PEOT/PBT, 300PEOT55PBT45, polyActive) with 300 Da polyethylene oxide and a PEOT/PBT weight ratio of 55/45 was acquired from polyVation B.V. (Groningen, the Netherlands). AM scaffolds were produced by means of screw-extrusion-based fused deposition modeling (FDM) (Bioscaffolder SYSENG, Germany). The FDM extrusion is controlled by the screw rotation and assisted by $N_2$ (5 bar) gas pressure allowing fine control over deposition of the molten polymer. The manufacturing of the 20 × 20 × 4 mm scaffolds was achieved following a layer-by-layer manufacturing with 90° rotation between deposited layers. The 3D scaffold CAD models were uploaded into PrimCAM software (Primus Data, Switzerland) and the deposition patterns were calculated. The fiber spacing, the distance between successive fibers in the same layer, was defined as 650 µm, the layer thickness was set to 170 µm, and the fiber diameter obtained was according to the nozzle diameter used, the polymer selected, and the processing parameters. The parameters that influence the production of the 3D scaffolds are temperature, screw rotation, deposition velocity. PEOT/PBT pellets were loaded in the reservoir and melted at a temperature of 195 °C or 220 °C, respectively. The screw rotation for the polymers was 200 rpm. The molten polymer was extruded through a nozzle with G25 (inner diameter = 250 µm). The deposition velocity was optimized to 200 mm min$^{-1}$. Figure S1b,c, Supporting Information shows the pore network architecture of the fabricated AM scaffolds.

*Cell Culture on 3D Scaffolds*: Preselected hMSCs (from a male aged 22) were retrieved from the Institute of Regenerative Medicine at Texas A&M University. All experiments were carried out with the full, informed consent of the subjects, in accordance with all local laws and with the approval of all relevant ethics bodies. Briefly, a bone marrow aspirate was drawn from the patient after informed written consent, and mononuclear cells were separated using density centrifugation. For expansion, hMSCs were cultured on tissue culture polystyrene plates at 1000 cells cm$^{-2}$ in basic medium, consisting of αMEM added with 10% fetal bovine serum (basic medium) (Thermo-Fisher Scientific), until 70–80% confluent. All experiments were performed with cells at passage 5. AM scaffolds, square blocks of 5 × 5 × 4 mm (width, length, height), were seeded with a drop of 35 µL containing 2 × 10$^5$ cells, unless stated otherwise. Two hours after seeding, the AM scaffold was flipped upside down to increase cell distribution. Four hours after seeding, scaffolds were transferred to a nontreated, 12-well plate for further culture. Scaffold cultures were performed in proliferation medium, consisting of basic medium supplemented with 1 ng mL$^{-1}$ fibroglast growth factor2 (Neuromics), 200 µM L-Ascorbic acid 2-phosphate (Sigma-Aldrich), and 100 U mL$^{-1}$ penicillin-streptomycin. Standard experimental conditions containing capsules for pH measurements were prepared by mixing cells with capsules right before seeding (Figure S2, Supporting Information). Here it is also important to mention that a higher proportion of capsules in the scaffold was toxic for the cells as it inhibited their adhesion to the scaffold and hampered cell culture. Hence, a dose-response experiment was conducted using FITC-capsules to determine the optimal capsule concentration balancing toxicity and spatial signal (data not shown). The capacity of hMSCs to attach to the scaffolds and to fill pores was evaluated by bright field microscopy, while the spatial capsule density was evaluated by epifluorescence collecting signal in the green channel ($\lambda$emission: 505–550 nm). Notably, it was concluded that the optimal seeding cell:capsule ratio to be used should be 1:1.75. Hence, considering these optimization results, about $3.5 \times 10^5$ capsules were added per scaffold. To this aim, a stock capsule suspension was diluted in 35 µL cell suspension containing $2 \times 10^5$ cells, obtaining a cell:capsule ratio of 1:1.75. Static cell cultures with capsules were prepared using a 35 µL containing $2 \times 10^5$ cells, as indicated above. The same proportion of cells and capsules per scaffold was maintained in dynamic seeding experiments. In particular, four scaffolds were placed in a 2 mL sterile Eppendorf tube along with four times the number of capsules and cells. The mixture of cell suspension and capsules was diluted in 1850 µL using $CO_2$-equilibrated proliferation medium. Cells in dynamic seeding condition were incubated on an orbital shaker at 30 rpm located in an incubator at 37 °C for seeding. Four hours after dynamic seeding, scaffolds were transferred to a nontreated, 12-well plate for further culture like static seeding experimental conditions. Media was changed every 72 h to minimize cell culture disturbance. Media refreshment was carried out moving scaffolds into 12-well plates filled with fresh and pre-equilibrated culture media (Figure S2, Supporting Information).

*Imaging*: Confocal laser scanning microscopy (CLSM) was performed with a Leica TCS SP5 (Leica Microsystems GmbH, Wetzlar, Germany) microscope. A Leica objective HCX APO L 20x/1.00 W was used for excitation and epicollection.

SNARF-1 was excited with 543 nm laser line and the emission signal was detected at 580–610 nm ($I_{595}$) and 625 – 655 nm ($I_{640}$). FITC was excited with 488 nm laser line, and the emission signal was collected was detected at 505–550 nm. LysoTracker Red DND-99 (L7528, ThermoFischer) was excited with 543 nm laser line and detected at 575–620 nm, calcein AM (C3100MP, ThermoFischer) was excited with 488 nm laser line and detected at 500–570 nm. CNA-35 (a kind gift from C. Reutelingsperger, Department of Biochemistry of Apoptosis, Maastricht University) was excited with 488 nm laser line and detected at 505–560 nm. Ethidium homodimer-1 (EthD-1, E1169, Thermo-Fischer) was excited with 514 nm laser line and detected at 590–640 nm. As a control, (PSS/PAH)$_3$ capsules, without SNARF-dextran into the cavities, were imaged under different CLSM settings (see Figure S3, Supporting Information) to ensure that the eventual polymer autofluorescence would not interfere with pH-sensing measurements.

For transmission electron microscopy (TEM), samples were prepared by fixation in 4% paraformaldehyde in phosphate-buffered saline, followed by washing with 0.1 M Cacodylate (3× for 15 min). Cells were fixed again with 2.5% glutaraldehyde in Cacodylate 0.1 M overnight (minimum of 1 h), followed by washing with 0.1 M Cacodylate (3× for 15 min), postfixed with 1% osmium tetroxide + 1.5% potassium hexacyanoferrate (II) trihydrate in Cacodylate 0.1 M, then washed again with 0.1 M Cacodylate 3× for 15 min. Then we proceeded to a dehydration series (70% for 30 min, 90% for 30 min, and two times 100% for 30 min), followed by propylenoxide 2× for 30 min and propylenoxide:Epon LX112 (1:1) overnight with stirring. Samples were covered with fresh epon LX112 for 7 h with stirring and embedded in beem capsules with fresh epon for 3 days at 60 °C. Sections (60 nm in size) were then cut with a diamond knife, stained with uranyl acetate and lead citrate, and imaged with a TEM (FEI Tecnai G2 Spirit BioTWIN iCorr).

*Sensing of pH in the 3D Cell Cultures Over Time*: Calibration of the pH sensors was performed by adding a small volume of capsules suspensions in Milli-Q water (5 µL) into pH-adjusted cell medium (30 µL). Final pHs were 5.64, 5.85, 6.1, 6.32, 6.46, 6.58, 6.7, 6.78, 6.89, 7.01, 7.13, 7.25, 7.4, 7.63, 8.01, and 8.28. Samples were mixed, allowed to equilibrate for 15 min, and then imaged by CLSM ($\lambda_{exc}$ 543 nm) to collect the fluorescence signals in the yellow (shown as false color in green, $\lambda_{em}$ 580–610 nm) and red channels ($\lambda_{em}$ 625–655 nm). At least four CLSM images were acquired for each pH point, with around 20–50 capsules in each image. Images were then processed with *ImageJ*.[54] Ratio was calculated by dividing $I_{595}$ image with $I_{640}$ image pixel by pixel. Before division, the eight-bit images were processed with the *Despeckle* filter from *ImageJ* to remove noise and thresholded to min and max values (min:20, max:250) to remove noise and saturated pixels. The ratio values for each pH fixed solution were used to fit a curve and calculate the calibration formula connecting ratio values with pH. Excel was used to fit the curve. Each point is the average of five measurements and error bars indicate the standard deviation.

The pH calibration formula is

$$pH = -0.6692R^3 + 3.154R^2 - 6.1577R + 11.103 \quad (1)$$

where $R$ is the emission intensity ratio ($I_{595\,nm}/I_{640\,nm}$).

*Statistical Significance*: The calibration data were acquired by combining results from independent experiments conducted over four different days. The comparison between pH at different locations within the scaffold, 2D, and aMEM was done using one-way analysis of variance (ANOVA) with Tukey's test for day 1 as well as day 7. $p$-value < 0.05 was used to determine the statistically significant difference between the means. The pH comparison between day 1 and day 7 for similar culture conditions was made using one-way ANOVA with Tukey's test. $p$-value < 0.05 was used to determine the statistically significant difference between the means. The sample size for each category was always ≥4 for the ANOVA.

## Supporting Information

Supporting Information is available from the Wiley Online Library or from the author.

## Acknowledgements

I.L.M. and A.C. contributed equally to this work. The authors are grateful to the European Community's Seventh Framework Program (FP7/2007-2013) (grant agreement No. 305436, STELLAR), the European Research Council (ERC) under the European Union's Horizon 2020 research and innovation programme (grant agreement No. 759959, INTERCELLMED) and the FISR/MIUR-C.N.R., Tecnopolo di Nanotecnologia e Fotonica per la medicina di precisione (project number: B83B17000010001) and Tecnopolo per la medicina di precisione - Regione Puglia (project number: B84I18000540002). The authors thank the M4I institute of the Maastricht University for their support with TEM. The authors further extent their acknowledgement to the Texas A&M Health Science Center College of Medicine Institute for Regenerative Medicine at Scott & White who isolated and provided the cells through a grant from NCRR of the NIH (Grant #P40RR017447).

## Conflict of Interest

The authors declare no conflict of interest.

## Keywords